\documentclass[]{aa}  

\usepackage{graphicx}
\usepackage{txfonts}
\usepackage{lipsum}
\usepackage{subcaption} 
\usepackage{lscape}             
\usepackage{placeins}           
\usepackage{mathtools}
\usepackage{amsmath} 
\usepackage[normalem]{ulem}
\usepackage{soul} 

\ifx\pdfoutput\undefined
\usepackage[hypertex]{hyperref}
\else
\usepackage[pdftex,hypertexnames=false]{hyperref}
\fi
\newcommand{\va}{\mathbf{a}}
\newcommand{\vx}{\mathbf{x}}
\newcommand{\vv}{\mathbf{v}}

\usepackage{xcolor}
\definecolor{myred}{HTML}{dc267f} 

\begin{document}

   \title{From particles to precision}

   \subtitle{Simulating subsonic turbulence with smoothed particle hydrodynamics}


   \author{Rubén M. Cabezón\inst{1,2}
          \and
          Domingo García-Senz\inst{3,4}
          \and
          Oliver Avril\inst{5}
          \and
          Osman Seckin Simsek\inst{6}
          \and
          Sebastian Keller\inst{7}
          \and          
          Axel S. Lechuga\inst{3}
          \and 
          Lucio Mayer\inst{8}
          \and
          Ralf Klessen\inst{5,9}
          \and
          Florina M. Ciorba\inst{6}
            }

    \institute{Center for Scientific Computing - sciCORE, University of Basel, Klingelbergstrasse 61, 4056 Basel, Switzerland\\
    \email{ruben.cabezon@unibas.ch}
    \and
    Center for Data Analytics - CeDA, University of Basel, Klingelbergstrasse 61, 4056 Basel, Switzerland
    \and
    Departament de Física. Universitat Politècnica de Catalunya (UPC). Av. Eduard Maristany 16, 08019 Barcelona, Spain\\
    \email{domingo.garcia@upc.edu}
    \and
    Institut d'Estudis Espacials de Catalunya (IEEC), 08860 Castelldefels (Barcelona), Spain 
    \and
    Universit\"{a}t Heidelberg, Zentrum f\"{u}r Astronomie, Institut f\"{u}r Theoretische Astrophysik, Albert-Ueberle-Str.\ 2, 69120 Heidelberg, Germany
    \and    
    Department of Mathematics and Computer Science, University of Basel, Spiegelgasse 1, 4051 Basel, Switzerland
    \and
    Swiss National Supercomputing Center (CSCS), Via Trevano 131, 6900 Lugano, Switzerland
    \and
    Department of Astrophysics, University of Zürich, Winterthurerstrasse 190, 8057 Zürich, Switzerland
    \and
    Universit\"{a}t Heidelberg, Interdisziplin\"{a}res Zentrum f\"{u}r Wissenschaftliches Rechnen, Im Neuenheimer Feld 225, 69120 Heidelberg, Germany
    }

   \date{Received September 30, 20XX}

 
  \abstract
   {The direct numerical simulation of subsonic turbulence with smoothed particle hydrodynamics (SPH) has traditionally been hampered by zeroth-order (E0) errors, inaccurate gradient evaluations, and excessive numerical dissipation.}
   {We aim to investigate whether a modern SPH formulation can overcome these challenges by comparing its results to those obtained with state-of-the-art non-SPH codes such as AREPO and GIZMO.}
   {We used the SPH-EXA code, a highly scalable, natively GPU-accelerated, state-of-the-art SPH code equipped with an integral-based gradient estimation, advanced artificial viscosity switches with a slope-limited reconstruction, a flexible family of pairing-resistant interpolation kernels, generalized volume elements, and improved handling of density contrasts, which maximizes Lagrangian compatibility.}
   {Our results show that SPH-EXA accurately reproduces the Kolmogorov inertial range scaling in the subsonic regime with increasing resolution, closely matching the results of state-of-the-art non-SPH methods. We also identify accurate grad-h terms as critical: a noisy standard implementation can imprint spurious granulation in the density field once dissipation is sufficiently reduced.}
   {We demonstrate that, with appropriate methodological advances, SPH can achieve a level of fidelity in modeling subsonic turbulence that rivals the most advanced Eulerian and moving-mesh approaches.}

   \keywords{hydrodynamics --
                turbulence --
                Methods: numerical
               }

   \maketitle
\nolinenumbers

\section{Introduction}
\label{sec:intro}
Smoothed particle hydrodynamics (SPH) is a mesh-free, conservative method for continuum mechanics widely used in many different fields since its conception around 1980 \citep{lucy_numerical_1977, gingold_smoothed_1977}. Its strengths include excellent conservation properties and a natural treatment of fully three-dimensional flows \citep{liu_smoothed_2010, price_smoothed_2012}. Despite these advantages, SPH has historically struggled in some regimes, most notably in accurately simulating subsonic turbulence.

Turbulence is a fundamental multiscale phenomenon in fluid dynamics, characterized by chaotic and nonlinear motion \citep{frisch_turbulence_1995}. 
It occurs in both natural environments and technological applications, making its study central across science and engineering. In astrophysics, turbulence influences processes ranging from accretion flows to the interstellar medium, shaping star formation, galaxy evolution, and the emergence of large-scale structure. Despite its ubiquity and importance, turbulence remains one of the most challenging phenomena to model and predict because of its complex, nonlinear nature. Progress in this field beyond the heuristic but very successful pioneering model of Kolmogorov \citep{kolmogorov_local_1941} can lead to advances in technology, improved environmental management, and deeper insight into fundamental processes of nature.

The direct numerical simulation (DNS) of turbulence requires high-resolution and state-of-the-art codes. In general, there are no significant discrepancies between SPH simulations and grid-based methods for supersonic turbulence \citep{kitsionas_algorithmic_2009}. Nevertheless, reproducing subsonic or transonic turbulence with SPH has traditionally been a challenge because it fails to produce results of sufficient quality to compete with grid-based methods for a similar resolution \citep{bauer_subsonic_2012}. The first studies of subsonic turbulence with SPH in the context of cosmological simulations \citep{dolag_turbulent_2005,valdarnini_impact_2011} found rough agreement with the Kolmogorov power spectra but only in a very limited range of wave numbers. 
More recent DNS attempts led to minor \citep{bauer_subsonic_2012} or moderate improvements \citep{price_resolving_2012}, with the latter highlighting the importance of reducing the viscosity away from shocks after showing that the Reynolds number is proportional to the Mach number in SPH simulations. 
However, the velocity power spectra still showed significant differences with those obtained with mesh-based calculations. 
All of these studies identified zeroth-order errors ($E0$), excessive dissipation, and inaccurate gradient estimation as the main causes of the difficulties in simulating turbulence with SPH in low-Mach numbers. 

In this work, we present SPH simulations of subsonic turbulence with SPH-EXA\footnote{\url{https://github.com/sphexa-org/sphexa}\\\url{https://deepwiki.com/sphexa-org/sphexa}} \citep{cavelan2020, keller2023}, a GPU-native, state-of-the-art, and highly scalable hydrodynamic code that incorporates the latest advances in High Performance Computing (HPC), such as efficient domain decomposition, GPU-based neighbor search, on-the-fly generation of initial conditions, and in situ visualization. It also incorporates the latest improvements to the SPH technique, such as an integral approach to calculating gradients, generalized volume elements, and a slope-limited reconstruction (SLR) of the velocity field to control dissipation, among other improvements (see Sect.~\ref{sec:ingredients} and \ref{sec:gradh}). Equipped with such a set of resources, we show that the simulations of subsonic turbulence with SPH-EXA can compete with those obtained using top reference codes such as AREPO \citep{springel_e_2010} and GIZMO \citep{hopkins_new_2015}. Our results show that, as resolution increases, we recover a growing portion of the inertial range of the Kolmogorov cascade. Owing to the excellent scaling properties of SPH-EXA, this will allow DNS to probe an ever larger fraction of the turbulent inertial range in the coming years.

The structure of the paper is as follows: Sect.~\ref{sec:ingredients} describes the key ingredients of our SPH formulation, and Sect.~\ref{sec:ICs} outlines the initial conditions for the simulations. We present the results in Sect.~\ref{sec:results}, and in Sect.~\ref{sec:summary} we summarize and discuss our findings. Finally, Appendix~\ref{app:highest_res} shows detailed images of our highest resolution simulation.
 

\section{SPH ingredients}
\label{sec:ingredients}
In this section, we discuss the ingredients of our SPH formulation that have the largest impact on our simulations of subsonic turbulence. Namely, the calculation of gradients, dissipation, and generalized volume elements. 
\subsection{Gradients}
\label{sec:grads}
The SPH-EXA code implements an integral approach (IA) to calculate derivatives \citep{garcia-senz_improving_2012,cabezon_testing_2012}.
The IA formalism calculates gradients of an arbitrary function, $f,$ by solving the following matrix equation:
\begin{equation}
\left[
\begin{array}{c}
\partial f/\partial x_1\\
\partial f/\partial x_2\\
\partial f/\partial x_3\\
\end{array}
\right]_a
=
\left[
\begin{array}{ccc}
\tau_{11} & \tau_{12} & \tau_{13} \\
   \tau_{21}&\tau_{22}&\tau_{23} \\
   \tau_{31}&\tau_{32}&\tau_{33}
\end{array}
\right]^{-1}_a
\left[
\begin{array}{c}
I_1\\
I_2\\
I_3\\
\end{array}
\right]_a\,,
\label{eq:IADmatrix}
\end{equation}

\noindent
where the $\tau$ tensor elements are defined as the weighted sum of the dyadic product of the relative distances $(\vx_b - \vx_a) \otimes (\vx_b - \vx_a)$:
\begin{equation}
\tau_{ij,a}=\sum_b V_b~(x_{i,b}-x_{i,a})(x_{j,b}-x_{j,a})~W_{ab}(h_a);~i,j=1,3\,.
\label{eq:tauijsph}
\end{equation}
\noindent
Here, $x_i$ is the vector position coordinates, $V_b$ the volume element, $W_{ab}$ the interpolation kernel, and $I_{i,a}$ the vector integral elements, which are calculated as
\begin{equation}
I_{i,a}=\left[\sum_b V_b f_b~(x_{i,b}-x_{i,a})~W_{ab}(h_a)\right];~i=1,3\,.
\label{eq:approxI}
\end{equation}

As a consequence, the gradient calculation is
\begin{equation}
\nabla_i f_a=\sum_b V_b (f_b-f_a) A_{i,ab}(h_a);~i=1,3\,,
\label{eq:modern_gradients}
\end{equation}

\noindent
where
\begin{equation}
A_{i,ab}(h_a)=\sum_{j=1}^3 c_{ij,a}(h_a)(x_{j,b}-x_{j,a}) W_{ab}(h_a)\,,
\label{eq:IAD_A_terms}
\end{equation}

\noindent
where the summation is up to three for 3D simulations, and $c_{ij,a}$ are the elements of the inverse of tensor $\mathcal{T}$; that is, $\mathcal{C}=\mathcal{T}^{-1}$ (Eq.~\ref{eq:IADmatrix}).

\subsection{Dissipation}
\label{sec:av}
The basic dissipation control in SPH-EXA was performed using the following prescription for artificial viscosity (AV):
\begin{equation}
\Pi_{ab}^{AV}=
\begin{cases}
-\frac{\alpha}{2}v_{ab}^{sig}~w_{ab} & \text{for~~$\vx_{ab}\cdot\vv_{ab} < 0$}\\
 0 & \text{otherwise}
\end{cases}
\label{eq:AV}
\,,\end{equation}

\noindent
where $\alpha$ is the viscous parameter and is set to one. The signal velocity between a pair of particles is estimated as follows:

\begin{equation}
    v^{sig}_{ab}=\frac{1}{2}(c_a+c_b)-2w_{ab}\,.
    \label{eq:vsignal}
\end{equation}
\noindent
Finally, $c_a$ and $c_b$ are the local speed of sound for each particle, and $w_{ab}=(\vv_{ab}\cdot\vx_{ab})/|\vx_{ab}|$.

A linear velocity field implies a smooth flow without the presence of shocks. Therefore, linear velocity fields should have a vanishing AV. We reconstructed the local velocity field for each particle by removing the linear component of the velocity jump that is used to trigger and calculate the AV in Eq.~\ref{eq:AV}. We refer to this procedure as slope-limited reconstruction (SLR) because the local velocity difference entering the AV term is first reconstructed from a linear estimate of the velocity field and then limited through a flow-dependent modulation\footnote{It should be noted that the SLR procedure was informally called  "Artificial Viscosity Cleaner (AVC)" in our arXiv-v1.}. In this sense, the method is closely related in spirit to the slope-limited reconstructions used in the Godunov methods, although we implemented it within a standard SPH AV framework, as originally suggested by \cite{frontiere_crksph_2017}. Assuming that it is differentiable in the neighborhood of each SPH particle, the best linear approximation of the velocity field is given by multiplying its Jacobian by the distance to its neighbors. Instead of using $\vv_{ab}=\vv_a-\vv_b$ to calculate $w_{ab}$, we used $\vv'_{ab}=\vv'_a-\vv'_b$, which is defined as
\begin{align}
    v'_{a,i}&\equiv v_{a,i}-\frac{1}{2}\phi_{ab}\mathbf{J}_{\vv_a}\vx^T_{ab} \label{eq:slrv1},\\
    v'_{b,i}&\equiv v_{b,i}+\frac{1}{2}\phi_{ba}\mathbf{J}_{\vv_b}\vx^T_{ab,} \label{eq:slrv2}
\end{align}

\noindent
where $\vx^T_{ab}=\vx_a-\vx_b$ is the column vector of the relative distance and $\mathbf{J}_{\vv}$ is the Jacobian of the velocity field. In the case of discontinuities, such a linear estimator is not accurate enough; hence, the $\phi_{ab}$ factor is designed to go down to zero in these cases, while it is kept close to one when the field is smooth. To this end, we used a van Leer-like limiter \citep{frontiere_crksph_2017}:

\begin{align}
    \phi_{ab} &=\max\left[0,\min\left(1,\frac{4F_{ab}}{(1+F_{ab})^2}\right)\right]\times\kappa_{ab}\,,\label{eq:phiab}\\
    \kappa_{ab} &\equiv
    \begin{dcases}
        \exp{\left[-\left(\frac{q_{ab}-q_{crit}}{q_{fold}}\right)^2\right]} & \text{if~~$q_{ab}<q_{crit}$}\,,\\
        1 & \text{if~~$q_{ab}\geq q_{crit}$}\,,
    \end{dcases}\label{eq:kappaab}\\
    q_{ab} &\equiv \min(q_a,q_b)\,,\\
    q_{crit} &\equiv \left(\frac{32\pi}{3nb_a}\right)^{1/3}\,,\\
    F_{ab} &\equiv \frac{\vx_{ab}\mathbf{J}_{\vv_a}\vx^T_{ab}}{\vx_{ab}\mathbf{J}_{\vv_b}\vx^T_{ab}}\,,
\end{align}

\noindent
where $q_{a}$ is the normalized distance between neighbors ($q_{a}=|\vx_b-\vx_a|/h_a$), $q_{crit}$ is the average interparticle distance normalized to the smoothing length in 3D, and $nb_a$ is the number of neighbors of particle $a$. The parameter $q_{fold}$ determines the width of the correction factor $\kappa_{ab}$ (Eq.~\ref{eq:kappaab}), which is applied to the van Leer limiter (Eq.~\ref{eq:phiab}). We used the recommended value \citep{frontiere_crksph_2017} of $q_{fold}=0.2$.

The final step is to control excessive dissipation effectively. This can be done by allowing for a dynamic individual viscous parameter of $\alpha_a$ in Eq.~\ref{eq:AV}, rather than having a constant value for all SPH particles. Several methods have been developed to calculate $\alpha_a$, but the nominal works of \citet{cullen_inviscid_2010} and \citet{read_sphs_2012} are the inspiration behind our implementation in SPH-EXA. The main idea is that $\alpha_a$ should be initialized by default at a rather low value (usually $\alpha_a=0.05$). It then increases rapidly if dissipation is needed and decays exponentially to its original value when it is not needed, effectively acting as a switch for the AV at the position of each SPH particle.

We started by calculating a local viscous parameter, $\alpha_{loc,a}$, following this definition \citep{read_sphs_2012}:

\begin{equation}
\alpha_{loc,a}=
\begin{dcases}
\frac{\alpha_{max}\mathcal{D}_a}{\mathcal{D}_a+h_a\vert\nabla\cdot \mathbf{v}_a\vert+0.05 c_a} & \text{for~~$\nabla\cdot\mathbf{v}_a < 0$}\,,\\
 0 & \text{otherwise}\,,
\end{dcases}
\label{eq:alphaloc}
\end{equation}

\noindent
where $\mathcal{D}_a=h^2_a\vert\nabla(\nabla\cdot\mathbf{v}_a)\vert$ predicts flow convergence by comparing the gradient of $\nabla\cdot\mathbf{v}_a$ with the local value of $\nabla\cdot\mathbf{v}_a$. If the former is large enough and the fluid is actually converging ($\nabla\cdot\mathbf{v}_a<0$), $\alpha_{loc,a}$ adopts a value between zero and $\alpha_{max}$. Furthermore, the term $0.05 c_a$, where $c_a$ stands for the local speed of sound, acts as a base value for the size of the velocity fluctuations that can trigger dissipation, and it also works as a term to prevent divergent $\alpha_{loc,a}$.

Once we knew $\alpha_{loc,a}$, we had two options:
\begin{itemize}
    \item if $\alpha_{loc,a}\geq\alpha_a$, we instantaneously increased the dissipation setting $\alpha_a=\alpha_{loc,a}$\,
    \item if $\alpha_{loc,a}<\alpha_a$, we let $\alpha_a$ decay smoothly\,
\end{itemize}

In the case where $\alpha_a$ had to decay, we evaluated an instantaneous decay rate of $\dot{\alpha}_a$ as in \citet{read_sphs_2012}:
\begin{equation}
    \dot{\alpha}_a=
    \begin{dcases}
        (\alpha_{loc,a}-\alpha_a)/\tau_a & \text{if~~$\alpha_{min}<\alpha_{loc,a}<\alpha_a$}\,,\\
        (\alpha_{min}-\alpha_a)/\tau_a & \text{if~~$\alpha_{min}\geq\alpha_{loc,a}$}\,.
    \end{dcases}
    \label{eq:alphadot}
\end{equation}

The decay time, $\tau_a$, is defined as $\tau_a=h_a/v_{sig,max}$, and it is a function of the local spatial resolution and the maximum signal velocity, $v_{sig,max}$, in the neighborhood of particle $a$, calculated according to Eq.~\ref{eq:vsignal}. The effect of Eq.~\ref{eq:alphadot} is that it allows alpha to decay exponentially to $\alpha_{loc,a}$, or $\alpha_{min}$ if $\alpha_{loc,a}$ is too low. This way, we ensured a minimum dissipation that helps reduce the random noise that might arise from particle disorder. However, we note that even though all particles have at least a minimum value of $\alpha_{min}$, dissipation is automatically cut off when $\mathbf{r}_{ab}\cdot\mathbf{v}_{ab}\geq0$, as stated in Eq.~\ref{eq:AV}. In addition, there is no restriction for $\alpha_{min}$, which can be set to zero.

In this respect, it is worth noting the latest work pointing to the interesting possibility of not needing switches at all, if a modulated version of the SLR is used \citep{price_switches_2024, sandnes2025, garcia-senz_SLR_2025}. This is achieved by taking a constant $\alpha=1$ and multiplying the limiter $\phi_{ab}$ in Eqs.~\ref{eq:slrv1} and \ref{eq:slrv2} by

\begin{equation}
    1-\mathcal{B}^p\,,
    \label{eq:SLR_modulation}
\end{equation}
\noindent
where $\mathcal{B}$ is the Balsara limiter
and
\begin{equation}
{\mathcal B}_{a}= \frac{\vert\nabla\cdot \mathbf v\vert_a}{\vert\nabla\cdot \mathbf v\vert_a + \vert \nabla\times \mathbf v\vert_a+10^{-4}\frac{c_a}{h_a}}\,.
\label{eq:balsara_1}
\end{equation}

\citet{garcia-senz_SLR_2025} showed that taking $p=2$ in Eq.~\ref{eq:SLR_modulation} provides the best results across a series of tests involving shocks and/or shear, significantly improving over the use of switches and switches+SLR techniques. However, in order to achieve the best results in simulating subsonic turbulence, the authors recommended including an additional control in the amplitude of the AV to avoid spurious triggering of dissipation by the random motion of particles in the smaller scales. This can be done with either a switch, as shown before, or directly with a limiter, such as Balsara.

\subsection{Volume elements}
\label{sec:ve}
In the context of SPH, where one discretizes the SPH integral interpolant, it becomes necessary to compute the volume element that corresponds to each particle. An easy yet highly effective method is $dx'^3\equiv V_b\sim m_b/\rho_b$. However, it is known \citep{saitoh_density-independent_2013,hopkins_general_2013} that the discretization of the volume element, present in most SPH equations, can be improved beyond this classical approximation using generalized volume elements (GVEs).

Generalized volume
elements are usually expressed as
\begin{equation}
    V_a=\frac{X_a}{\sum_{b=1}^{nb}X_bW_{ab}}\,,
    \label{eq:sph_gve_def}
\end{equation}
\noindent
where $X$ is an arbitrary function. In SPH-EXA, we adopted the GVE described in the work of \citet{garcia-senz_conservative_2022}, which has been shown to provide a better treatment of discontinuities, stronger resistance to tensile instability, better partition of unity, and more accurate gradient evaluation:

\begin{equation}
    X_a=\frac{m_a}{\rho^0_a}\,,
    \label{eq:sph_gve}
\end{equation}
\noindent
where $\rho^0_a=\sum_bm_bW_{ab}$ is the standard SPH density calculation. With Eqs.~\ref{eq:sph_gve_def} and \ref{eq:sph_gve} we can simply calculate the particle density as $\rho_a=m_a/V_a$.

\subsection{SPH-EXA hydrodynamic equations}
\label{sec:equations}
The final set of hydrodynamic equations in SPH-EXA is as follows:

\begin{align}
  \rho_a = &m_a/V_a\,,\\
  \begin{split}
  \frac{dv_{i,a}}{dt} = &- \sum_b m_b\left[\frac{X_a^{2-\sigma}X_b^{\sigma} P_a}{\Omega_a m_a^2~k_a} A_{i,ab}(h_a) +\frac{X_b^{2-\sigma}X_a^{\sigma} P_b}{\Omega_b m_b^2~k_b} A_{i,ab}(h_b)\right]\,,\label{eq:mom}
  \end{split}\\
  \begin{split}
  \frac{du_a}{dt} =& \frac{X_a^{2-\sigma} P_a}{m_a^2\Omega_a k_a}\sum_b\sum^d_{i=1} m_b X_b^{\sigma}\left[(\vv_{a}-\vv_{b})\cdot \mathbf{A}_{ab}(h_a)\right]+\left(\frac{du_a}{dt}\right)^{AV}\,.\label{eq:ener}
  \end{split}
\end{align} 

\noindent
Here, $P_a$ is the pressure, $X_a=m_a/\rho^0_a$ the selected function for the GVE, $k_a$ the normalization of the GVE ($k_a\equiv\sum_bX_bW_{ab}$ in Eq.~\ref{eq:sph_gve_def}), $A_{i,ab}(h_a)$ the IA term, and $\Omega_a$ the grad-h term. The parameter $\sigma$ controls how close the implementation is to a pure Lagrangian scheme \citep[see][]{garcia-senz_conservative_2022}. The last term in the energy equation includes the contributions of AV.

\section{Initial conditions}
\label{sec:ICs}
We designed an embedded generator of initial conditions (ICs) in SPH-EXA. In all our tests, we used a template of a relaxed box with $50^3$~particles. This box has previously been relaxed to achieve a glass-like configuration \citep{arth2019} with constant density and periodic boundary conditions. Replicating this box in different directions allowed us to create on-the-fly ICs with a user-defined number of particles in a distributed manner; this was only limited by the memory of the system.

All simulations used a periodic box of $L=1$ in size with a uniform initial density of $\rho=1,$  pressure of $P=1$, and a quasi-isothermal equation of state ($\gamma=1.001$). Turbulence is driven by stochastic forcing modes at different wavelengths, \textbf{k,} in the $1<|\mathbf{k}|<3$ range, with amplitudes following a parabolic distribution of $\propto 1-(k-k_c)^2$ peaking at $k_c=2$, ensuring that kinetic energy is injected only at the largest scales of the spectrum. The resulting forcing is built from a mixture of $1/3$ compressive and $2/3$ solenoidal modes, whose phases evolve following a stochastic Ornstein-Uhlenbeck process \citep{ornstein_brownian_1930,federrath_statistics_2010}. Simulations start from homogeneous ICs and are driven for $t=10$ (in units of sound-crossing time, $t_{sc}=L/c=1$), until a statistically stationary state with an RMS Mach number of $\mathcal{M}\sim0.34$ is reached.

\section{Results}
\label{sec:results}

\subsection{Impact of solver elements at fixed resolution}
\label{sec:results_elements}
\begin{figure}[t]
    \centering
    \includegraphics[width=\linewidth]{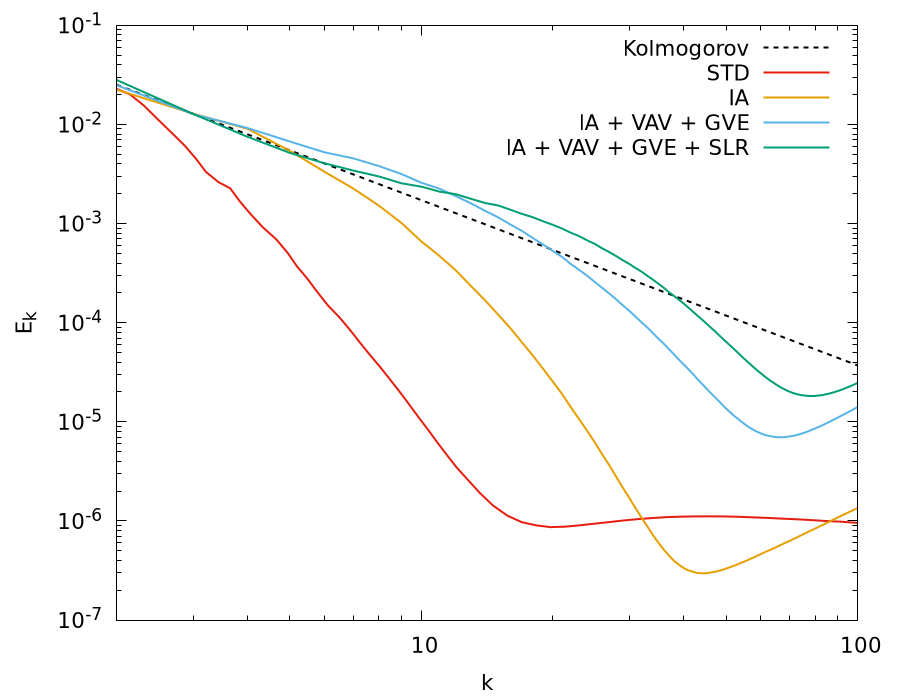}
    \caption{Comparison of the power spectra of the velocity field with different elements incorporated into the SPH solver. The dashed line is the theoretical Kolmogorov power spectrum of $\sim k^{-5/3}$. STD denotes standard SPH and is taken from \citep[simulation S3 in their Fig.~6, with $256^3$ particles]{bauer_subsonic_2012}. All other simulations were performed with SPH-EXA and $250^3$ particles. IA stands for including the integral formalism to calculate derivatives. VAV stands for variable AV and implies the inclusion of AV switches. GVE is for generalized volume elements, and SLR for slope-limited reconstruction. All these elements are also described in Sect.~\ref{sec:ingredients}.} 
    \label{fig:spectra_elem_comp}
\end{figure}

To evaluate the weight and significance of the different elements that contribute to our SPH implementation, we conducted identical simulations while progressively incorporating more of these elements. Figure~\ref{fig:spectra_elem_comp} shows the power spectrum of the velocity field of simulations of $250^3$ particles with increasing complexity, compared to the S3 simulation of \citet{bauer_subsonic_2012} with $256^3$ particles and the classic SPH formulation (i.e., gradients calculated directly as kernel derivatives, with no GVEs and no viscosity switches); this is labeled as STD in our plot.

\begin{figure}[t]
    \centering
    \includegraphics[width=\linewidth]{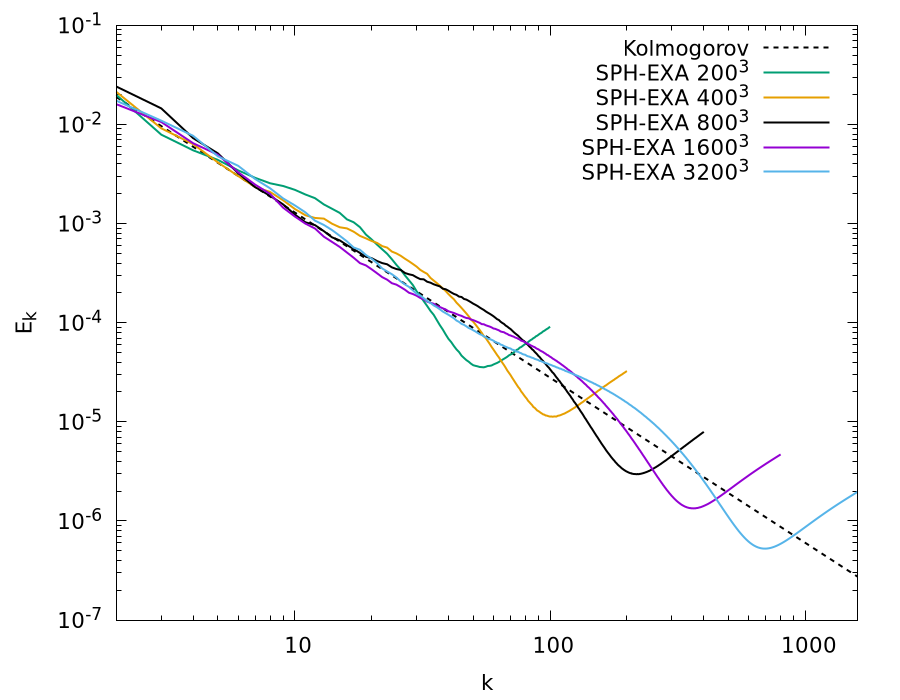}
    \caption{Comparison of instantaneous power spectra of the velocity field at $t=10$ (in units of sound-crossing time: $t_{sc}=L/c$) at increasing resolution. Each line corresponds to an increase of a factor of two in resolution with respect to the previous curve.}
    \label{fig:spectra_comp}
\end{figure}

\begin{figure*}
    \centering
    \includegraphics[width=\textwidth]{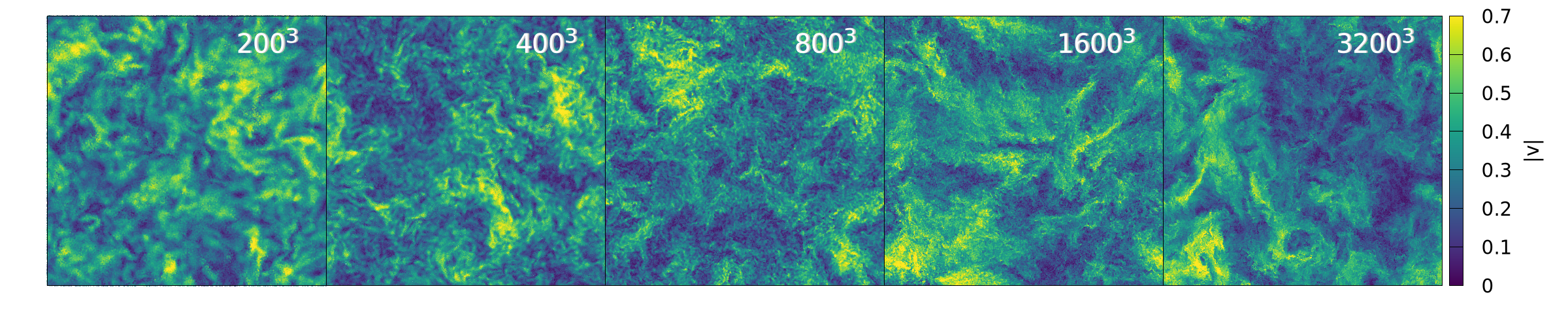}
    \caption{Slice of thickness $2h$, centered on $z=0$, showing the velocity magnitude with increasing resolution (SPH-EXA). We represent particles directly.}
    \label{fig:vmod}
\end{figure*}

The inclusion of IA recovers the correct slope in the inertial range of the Kolmogorov cascade at low $k$, considerably extending it compared to STD  \citep{valdarnini_improved_2016}, but it does not remove the overly steep slope $\sim -4$ at large $k$. At these small scales something else is needed. Adding GVE and especially AV switches (labeled VAV) moves in the right direction, revealing the bottleneck effect as an energy excess near the resolution limit, as commonly seen in mesh codes. Finally, the slope-limited reconstruction (SLR) provides the decisive improvement: it extends the inertial range by almost a decade relative to STD, makes the bottleneck clearly visible, and yields a dissipative-range slope comparable to non-Lagrangian codes (see Fig.~\ref{fig:spectra_comp_codes}).

\subsection{Resolution scaling and flow morphology}
\label{sec:results_resolution}
\begin{figure}
    \centering
    \includegraphics[width=\linewidth]{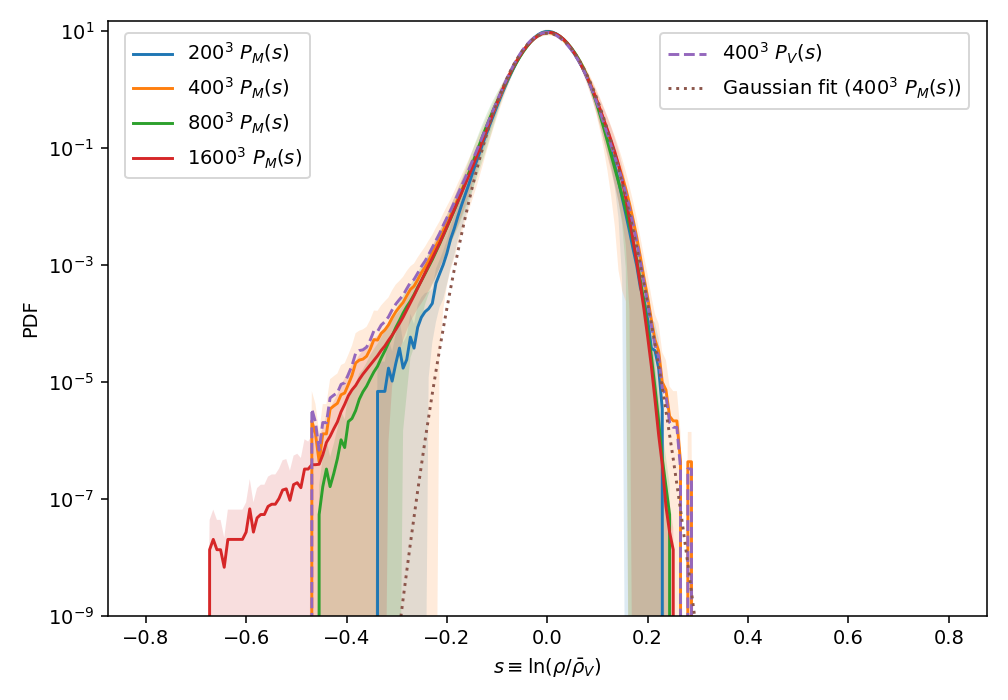}
    \caption{Density PDF for different resolutions. Solid lines show the averaged mass-weighted PDFs. Dashed and dotted lines show the averaged volume-weighted PDF and the Gaussian fit for $400^3$, respectively.}
    \label{fig:pdf}
\end{figure}

In Fig.~\ref{fig:spectra_comp} we present the power spectrum of the velocity field for simulations performed with SPH-EXA and with different particle counts at the end of each simulation ($t=10$). It is clear that with increasing resolution, the inertial range over the Kolmogorov scale is extended to a larger wave number, and, more importantly, that the rate of enhancement with increasing number of particles is not marginal or small, but considerable and consistent with that obtained with modern mesh-based and moving-mesh methods. This is a novel result, which means that SPH can indeed resolve smaller scales with higher resolution, in contrast to the results found with older SPH implementations  \citep[e.g.,][]{bauer_subsonic_2012}. Figure~\ref{fig:vmod} shows a thin slice of the velocity field for increasing resolution with SPH-EXA, from $200^3$ up to $3200^3$, where it is evident how we can resolve a delicate and rich mixture of large- and small-scale features.
Figure~\ref{fig:pdf} shows the probability density function (PDF) for particle density at different resolutions, averaged over five snapshots (at $t=10,15,20,25,30$). Shaded regions show the $\pm1\sigma$ scatter, and the dashed line represents the volume-weighted PDF for the $400^3$ simulation. The bulk of the PDF is robust across resolutions and close to Gaussian in its central part, while the tails, especially at low-density, show some resolution dependence, likely due to a slower convergence of intermittent events. We can also see the expected tail bias toward low densities, since in subsonic flows, local pressure increases in colliding flows can prevent the buildup of high-density regions \citep{konstandin2012}. 

\subsection{Cross-code comparison}
\label{sec:results_codes}
\begin{figure}
    \centering
    \includegraphics[width=\linewidth]{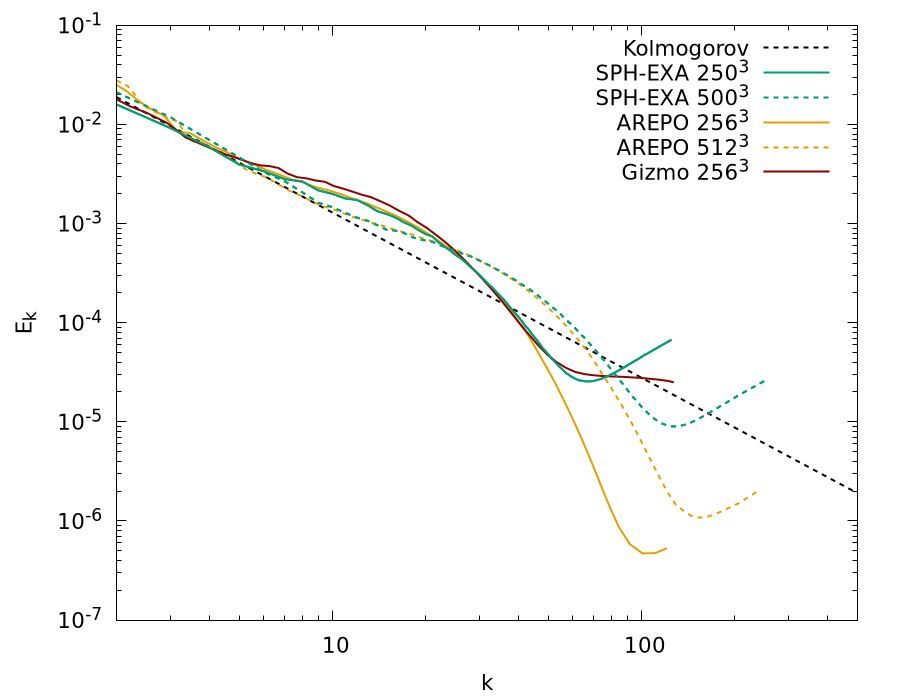}
    \caption{Comparison of velocity power spectra for different codes at $t=10$. AREPO and GIZMO curves are from their respective papers \citep{bauer_subsonic_2012, hopkins_new_2015}.}
    \label{fig:spectra_comp_codes}
\end{figure}

We can now compare our results directly to other state-of-the-art codes: moving-mesh and meshless finite-volume methods. Figure~\ref{fig:spectra_comp_codes} shows velocity power spectra from AREPO, GIZMO, and SPH-EXA. All three reproduce a comparable inertial range and a bottleneck of similar magnitude at comparable $k$. This suggests that once the injection method and the effective viscosity are matched, the cascade between the forcing and dissipation wave numbers behaves similarly across methods. We note, however, that we compare runs with a similar number of fluid elements, which does not necessarily imply the same spatial resolution.

In SPH, resolution depends on the shape of the entire kernel, not just on its compact support (usually $2h$), which is a crude upper bound. The smoothing length ($h$ instead of $2h$) is already a better proxy for local spatial resolution, since the largest contributions to the interpolations come within this scale. However, a more kernel-aware measure is to use the full width at half maximum (FWHM) of the kernel \citep{cabezon_mixedsincs_2024}. In SPH-EXA we used the sinc family of kernels \citep{cabezon_sinc_2008} with exponent $n=6$:
\begin{equation}
    W(h_a,q_a)=\frac{K_{3D}}{h_a^3}
    \begin{cases}
    1& \text{if }q_a=0\\
    \left[\frac{\sin(\frac{\pi}{2}q_a)}{\frac{\pi}{2}q_a}\right]^6& \text{if }0<q_a\le2\\
    0&\text{if }q_a>2
    \end{cases}
,\end{equation}

\noindent
where $K_{3D}$ is the normalization constant in 3D. The FWHM is defined as $W(h_a,q_{1/2})=\frac{1}{2}W(h_a,0)$, which gives $q_{1/2}\simeq0.5239$. Given a compact support of $2h$, the corresponding FWHM in units of smoothing length is
\begin{equation}
    \ell_{FWHM}=2hq_{1/2}\simeq1.05h    
.\end{equation}

In our $400^3$ simulation, the mean smoothing length is $\left<h\right>\sim 3.6\times 10^{-3}$, giving $\left<\ell_{FWHM}\right>=3.78\times 10^{-3}$. This is close to the approximate cell size in the AREPO run with $256^3$ cells $\sim 1/256\sim 3.9\times 10^{-3}$. The close agreement of the spectra at similar fluid element count, despite the lack of a one-to-one match between SPH and grid-based spatial resolution proxies, indicates that, in subsonic turbulence, the effective spectral resolution is very much controlled by gradient accuracy and numerical dissipation rather than just by the nominal smoothing/cell scale. Spatial resolution matters the most in the dissipation regime, where AREPO, GIZMO, and SPH-EXA have different transfer functions. In particular, GIZMO and SPH-EXA show a similar gentler slope in the roll-off, while AREPO is slightly steeper, pointing to a stronger high-$k$ damping. However, these differences are modest and the qualitative behavior persists at higher resolution, with the break shifting to larger $k$, as expected.

\begin{figure}
    \centering
    \includegraphics[width=\columnwidth]{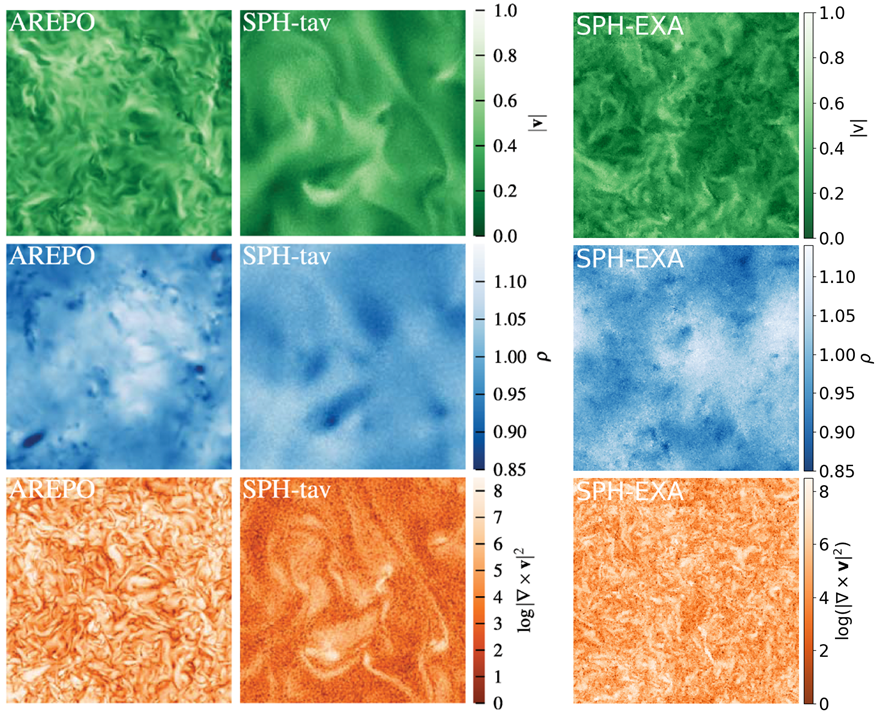}
    \caption{Comparison of SPH-EXA results with AREPO and SPH-tav in \citet{bauer_subsonic_2012} at $t=25$. From top to bottom: Velocity magnitude, density, and enstrophy. The two columns on the left are an excerpt of the original results of AREPO and their SPH with reduced AV (both with $256^3$ fluid elements) of \citet{bauer_subsonic_2012} shown in their Fig.~4. The column on the right is the result of SPH-EXA with $250^3$ particles. The fields show particles in a thin slice around $Z=0$. No rasterization was performed.}
    \label{fig:comparisonBauer}
\end{figure}

Figure~\ref{fig:comparisonBauer} shows, from top to bottom, the velocity magnitude, density, and enstrophy ($|\nabla\times\vv|^2$) of the fluid at the end of our simulations. The two columns on the left are an excerpt of the results with AREPO and the SPH used in \citet{bauer_subsonic_2012}. Since their plots are at $t=25$, we also extended our calculations to match the same time for this comparison. It is clear from these two columns that the SPH used in that work was unable to resolve subsonic turbulence because it lacks all small-scale features. The column on the right shows the results of SPH-EXA with $250^3$ particles. The similarities with the results of AREPO are evident, and there is a rich mix of fine structures. This can also be seen in Fig.~\ref{fig:vmod}, which illustrates a slice of the velocity magnitude at $t=10$ for different resolutions. SPH-EXA also produces an excellent second-order velocity structure function (Sect.~\ref{sec:sf2}). Additional color maps of the highest resolution simulation, $3200^3$, can be found in Appendix~\ref{app:highest_res}. 

\subsection{The relevance of accurate grad-h terms}
\label{sec:gradh}

\begin{figure}
    \centering
    \includegraphics[width=\linewidth]{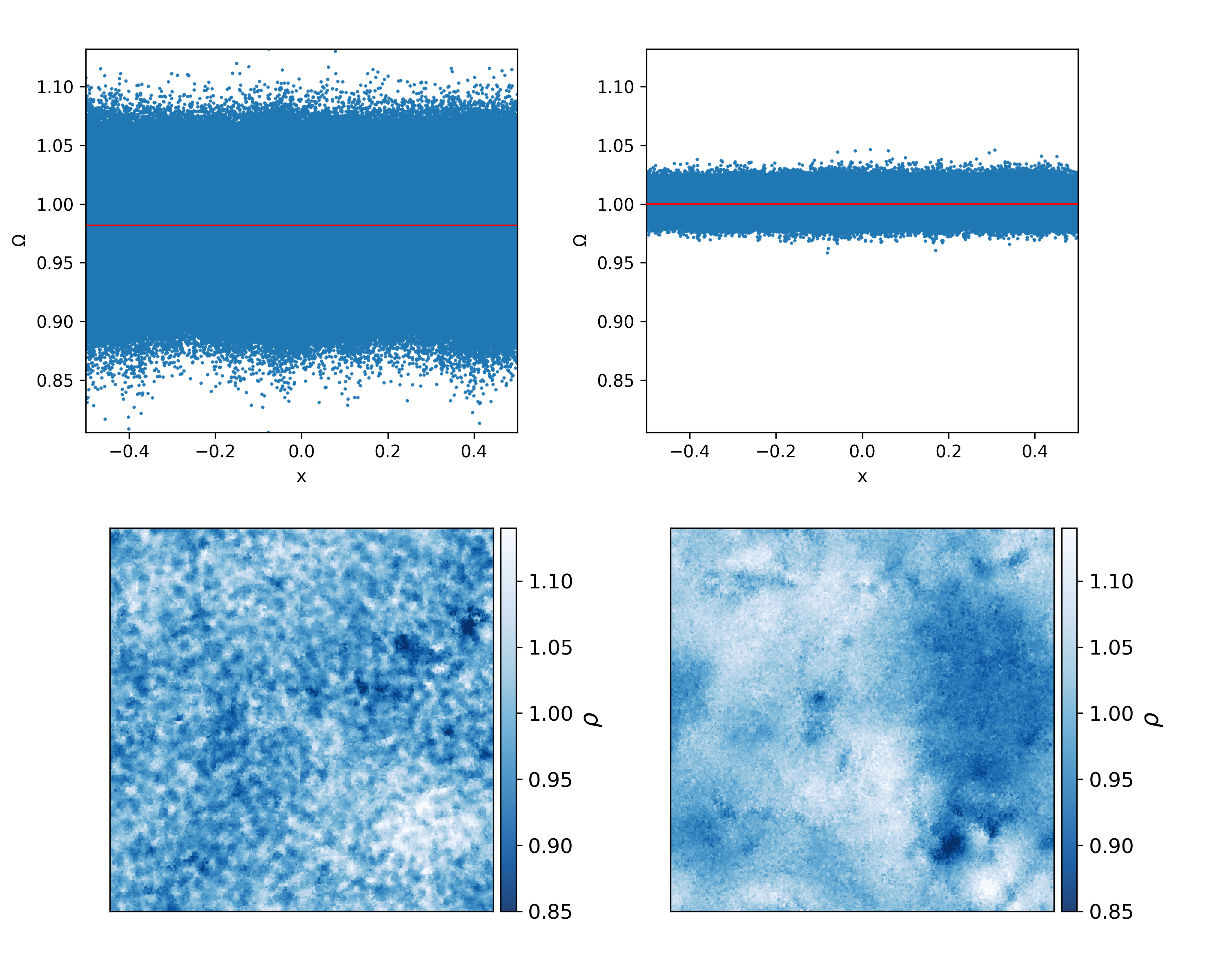}
    \caption{Effect of accuracy in grad-h terms. The top row shows the value of the grad-h correction, $\Omega,$ for all particles in a $250^3$ simulation at $t=10$. Left: $\Omega$ is calculated using the density field directly. Right:  $\Omega$ is calculated using the differences in the density field. The red lines show the mean value of $\Omega$. The bottom row shows a thin slice of the resulting density field.}
    \label{fig:granulation}
\end{figure}

Grad-h terms are an important addition to the SPH technique when using variable smoothing lengths ($h$) \citep{springel2002, monaghan2002}. It takes into account the spatial variation of $h$ as a corrective term in the momentum and energy equations (Eqs.~\ref{eq:mom} and~\ref{eq:ener}):
\begin{equation}
    \Omega_a = \left[1+\frac{h_a}{3\rho_a}\frac{\partial \rho_a}{\partial h_a} \right]^{-1}\,.
    \label{eq:gradh}
\end{equation}These corrections are a standard method to improve accuracy and consistency in particle interpolation, reducing numerical instability in variable-density simulations, and as such, they are included by default in most astrophysics SPH codes. 

During the subsonic simulations presented above, we noticed that the density field only appeared "granulated" when grad-h corrections were used jointly with SLR dissipation (see Fig.~\ref{fig:granulation}, left column). Initially, we suspected that the new addition (the SLR dissipation control) was the source of the problem, but it turned out that the problematic term was how the grad-h correction was calculated. After inspecting the values of this correction, we noticed that the average was systematically below $1$ and with a relatively large spread. For example, for the $250^3$ simulation, $\left<\Omega\right> =0.98 \pm 0.16$. This does not seem like much, but it is exactly the kind of "small" inconsistency that becomes relevant in subsonic turbulence, in particular when the dissipation formulation is designed to \textit{\emph{stop the damping of that noise}}.

In low-Mach turbulence, physical density fluctuations are weak, so any residual inconsistency in the pressure–density–$h$ coupling becomes visible. We can write the grad-h field as

\begin{equation}
    \Omega_a=\bar{\Omega}+\delta\Omega_a\,.
\end{equation}Then, we expand the multiplicative term in the SPH equations:

\begin{equation}
    \frac{1}{\Omega_a}=\frac{1}{\bar{\Omega}}\left(1- \frac{\delta\Omega_a}{\bar{\Omega}}+\mathcal{O}(\delta\Omega_a^2)\right)
.\end{equation}

The multiplicative factor, $1/\Omega_a$, that scales the pressure forces has two error sources: a biased mean and a noisy scatter. The first is just a constant scaling that affects the effective compressibility and slightly shifts the global balance. It shows as a systematic drift: a slightly different RMS Mach, a different effective forcing response, and slightly different spectral cut-off location, but it does not introduce particle-scale structure by itself. The second is the actual source of the granularity, since it is a multiplicative noise field coupled to the pressure acceleration. Since $\delta\Omega_a$ varies on the inter-particle scale and correlates with local disorder, it acts as a high-$k$ forcing of the compressive jitter:

\begin{equation}
    \va_a^P\approx\frac{1}{\bar{\Omega}}\va^P_{\Omega=1}-\frac{1}{\bar{\Omega}^2}\delta\Omega_a\va^P_{\Omega=1}+\dots
.\end{equation}If these corrections are in fact a noisy field, they induce noisy accelerations and particle jitter. This leads to noisy density estimates, noisy smoothing length, and then goes back to noisy grad-h corrections, creating a feedback loop that distorts the field completely. This is usually masked by baseline AV, but it becomes apparent when SLR suppresses dissipation in smooth regions. 

According to Eq.~\ref{eq:gradh}, we needed to evaluate $\partial  \rho_a/\partial h_a$, and we caution that the direct derivative with respect to $h$ of the density field can be too inaccurate. This is well known to have considerable E0 errors, and a typical trick is to include a scalar in the derivative to calculate it using the differences of the field, rather than the field directly. This has been used extensively to yield built-in conservation in SPH and zero derivatives when the field is constant. For example:
\begin{equation}
    \frac{\partial \rho}{\partial h}=\frac{1}{\phi}\left[\frac{\partial(\phi\rho)}{\partial h}-\rho\frac{\partial\phi}{\partial h}\right]\,,
\end{equation}
\noindent
where $\phi$ could be any other field or scalar. If we simply choose $\phi=1$,

\begin{equation}
    \frac{\partial \rho_a}{\partial h_a}=\sum_b V_b \rho_b \frac{\partial W_{ab}(h_a)}{\partial h_a}-\rho_a\sum_b V_b \frac{\partial W_{ab}(h_a)}{\partial h_a}\,.
    \label{eq:gradhcorrection}
\end{equation}The first term on the RHS of Eq.~\ref{eq:gradhcorrection} is the common grad-h term that is used to calculate $\Omega_a$. The second term on the RHS of Eq.~\ref{eq:gradhcorrection} is the additional correction that allows the grad-h term to be computed from pairwise density differences rather than from the standard formulation, substantially reducing its noise.

The practical implementation of this term depends on the volume elements used. In SPH-EXA, we corrected the calculation of $\partial \rho_a/\partial h_a$, amending formula A17 from \citet{garcia-senz_conservative_2022}. Defining

\begin{equation}
    \Psi_a\equiv \rho_a\sum_b V_b \frac{\partial W_{ab}(h_a)}{\partial h_a}\,,
\end{equation}

\noindent
we obtain
\begin{align}
    \begin{split}
    \frac{\partial\rho_a}{\partial h_a}=\left[\frac{\rho_a}{\rho^0_a}-X_a W_{aa}(h_a)\right]&\sum_b m_b \frac{\partial W_{ab}(h_a)}{\partial h_a}-\rho^0_a\Psi_a+\\
    \frac{m_a}{X_a}&\sum_b X_b \frac{\partial W_{ab}(h_a)}{\partial h_a}-\rho_a\Psi_a\,.
    \end{split}
\end{align}
\noindent
Note that if this formulation is used, the self-contributions of the particle $a$ must not be included. There are different ways to implement this correction, as well as different possibilities for the field $\phi$. We opted for the easiest and least invasive methodology.
As a result, Fig.~\ref{fig:granulation} (right column) shows that the average value of $\Omega$ improved considerably, becoming $\left<\Omega\right>=1.00 \pm 0.04$, and the density field was recovered. All production results reported in this work use the corrected grad-h formulation, and the uncorrected version is shown only in Fig.~\ref{fig:granulation} for diagnostic purposes.


\subsection{Second-order velocity structure function}
\label{sec:sf2}

\begin{figure}
    \centering
    \includegraphics[width=\linewidth]{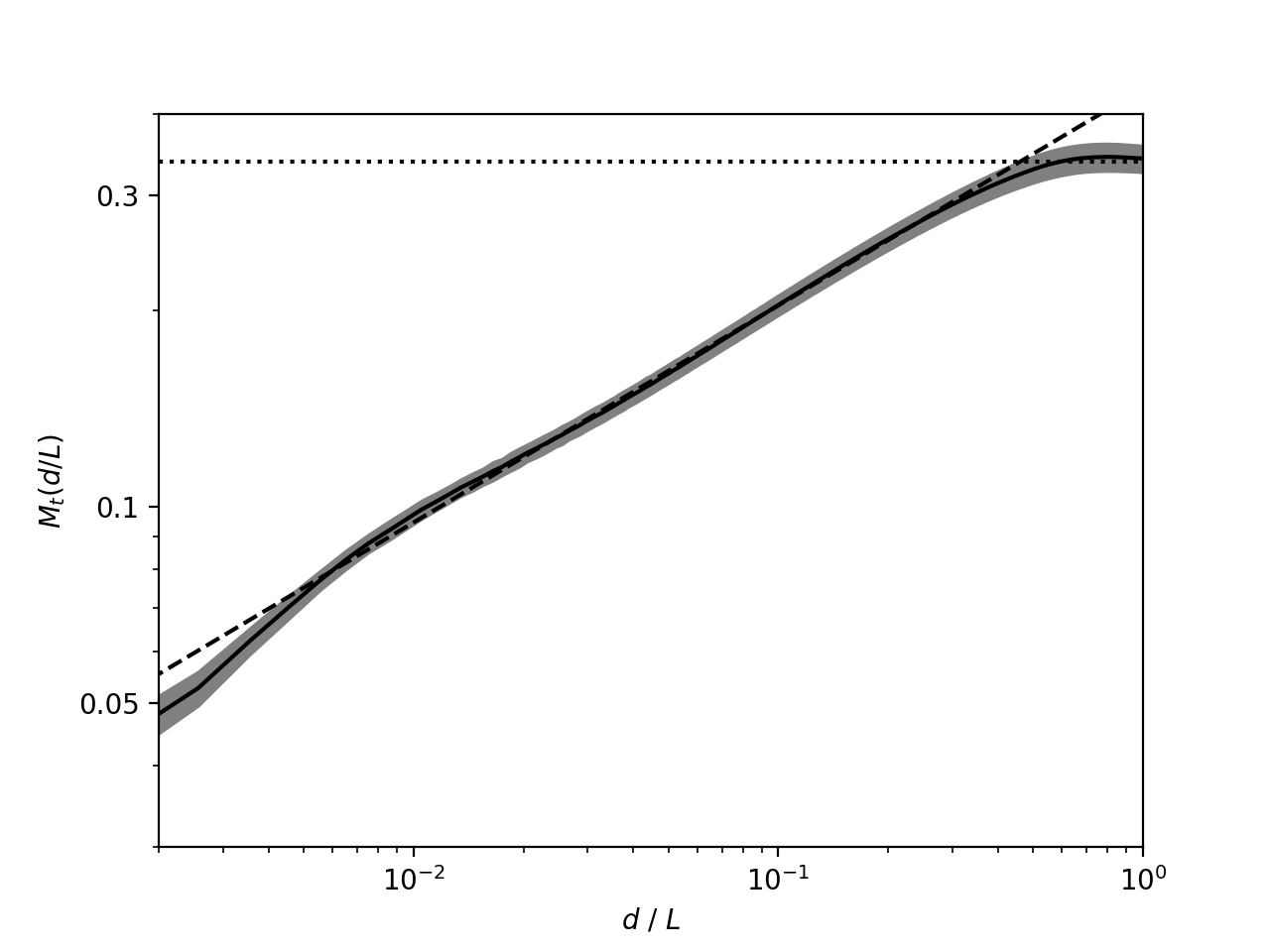}
    \caption{Mass-weighted turbulent Mach number as a function of spatial scale for a turbulent box simulation with a resolution of $2000^3$ particles. The dotted line indicates the RMS velocity of the simulation. The dashed line is a power-law fit to the turbulent cascade region.}
    \label{fig:structure-function}
\end{figure}

The SPH method is particularly advantageous for a wide range of mass-weighted analysis methods. Most importantly, it eliminates the need for special tracer particles, as the SPH particles themselves serve as tracers. This allows for an easy computation of several quantities such as the finite-time Lyapunov exponent and mass-weighted structure functions. The structure function is defined as 
\begin{equation}
    {SF}_2 (d/L) = \left \langle | \vec{v}(\vec{r}) - \vec{v}(\vec{r} + \vec{d}/L) |^2 \right \rangle_{\vec{r}}
,\end{equation}
\noindent
where $d=|\vec{d}(\vec{r})|$ is the distance between the two particles in a pair, and we compute the average over many SPH particle pairs, making this a mass-weighted structure function. $L$ is the turbulent driving scale. From this, one can compute the turbulent Mach number as 
\begin{equation}
    M_t (d/L) = \sqrt{{SF}_2 (d/L) / (2 c_s^2)}
.\end{equation}

As an example, Figure~\ref{fig:structure-function} shows the turbulent Mach number as a function of the spatial scale for one of our SPH-EXA simulations. The calculation was performed using $10^9$ particle pairs per snapshot in a $2000^3$ simulation and was averaged over $41$ nonconsecutive snapshots, which is sufficient to achieve statistical convergence. This resolution was chosen because it provided a statistically converged structure function at a manageable computational cost. A higher resolution run would yield a larger inertial range, but it would not significantly change the fitted slope.

In the regime of the turbulent cascade, we find a power law with an exponent of $0.333\pm0.001$. This matches the expected result of $1/3$ for subsonic Kolmogorov turbulence. Together with the density PDF shown in Figure~\ref{fig:pdf}, this demonstrates that SPH-EXA is capable of producing accurate mass-weighted statistics for subsonic turbulence in a much more natural way than comparable mesh codes.

\section{Summary and conclusions}
\label{sec:summary}
SPH-EXA combines a modern SPH formulation with state-of-the-art parallelization and GPU acceleration. This enables simulations that mitigate historical SPH limitations, such as excessive dissipation and inaccurate gradient evaluations, which have long hindered SPH from matching the performance of grid-based and moving-mesh codes in this field. Our findings not only align well with results from non-SPH codes such as AREPO and GIZMO, they also illustrate the robustness of SPH in handling complex turbulent flows at resolutions previously thought unattainable with this method. 

We demonstrate that collectively incorporating various components that address problems at multiple simulation scales achieves these results. Isolated enhancements in the SPH framework cannot tackle subsonic turbulence; however, a synergy of these improvements can. In particular, our combination of an integral approach to derivatives, generalized volume elements, artificial viscosity switches, accurate grad-h corrections, and slope-limited reconstruction shows excellent results.

A particularly relevant key finding is that noisy implementations of grad-h terms act as a multiplicative force error that becomes visible in low-Mach turbulence once the baseline dissipation is suppressed. In this respect, we propose an implementation based on the differences of the grad-h field, rather than on its absolute value. This yields substantially more accurate results in the calculation of the grad-h terms and, as a consequence, in the simulation of subsonic turbulence.

Finally, we demonstrate that, with increasing resolution, the inertial range of the power spectra extends to larger wave numbers, exhibiting features comparable to those seen in non-SPH codes. These improvements allow SPH to better exploit its intrinsic strengths: truly Lagrangian sampling (particles are the fluid elements, avoiding tracer advection and particle-to-grid interpolation noise), excellent conservation properties (exact mass advection and pairwise-conservative exchanges of momentum and energy), and flexibility across physical regimes. In particular, SPH couples naturally and accurately to tree methods for self-gravity, supporting its use in more demanding astrophysical contexts where turbulence is relevant.

\begin{acknowledgements}

This work has been supported by the Swiss Platform for Advanced Scientific Computing (PASC) project "SPH-EXA: Optimizing Smoothed Particle Hydrodynamics for Exascale Computing". It has also been carried out as part of the
SKACH consortium through funding from SERI.  We also acknowledge the financial support of the European Research Council through the ERC Synergy Grant ``ECOGAL'' (project ID 855130),  from the German Excellence Strategy via the Heidelberg Cluster of Excellence (EXC 2181 - 390900948) ``STRUCTURES'', and from the German Ministry for Economic Affairs and Climate Action in project ``MAINN'' (funding ID 50OO2206). The numerical calculations have been supported by the EuroHPC-JU Extreme Scale Access Mode through the project "TGSF: The role of turbulence and gravity in star formation" (EHPC-EXT-2023E01-031). The authors acknowledge the support of the LUMI Supercomputer Center, the Swiss National Supercomputing Center (CSCS - allocations c32 and ch18), and the Center for Scientific Computing (sciCORE) at the University of Basel.

\end{acknowledgements}
\bibliographystyle{aa}
\bibliography{bibliography}

\begin{appendix}
\section{High-resolution view}
\label{app:highest_res}

\begin{figure}[bh!]
    \centering
    \includegraphics[width=\linewidth]{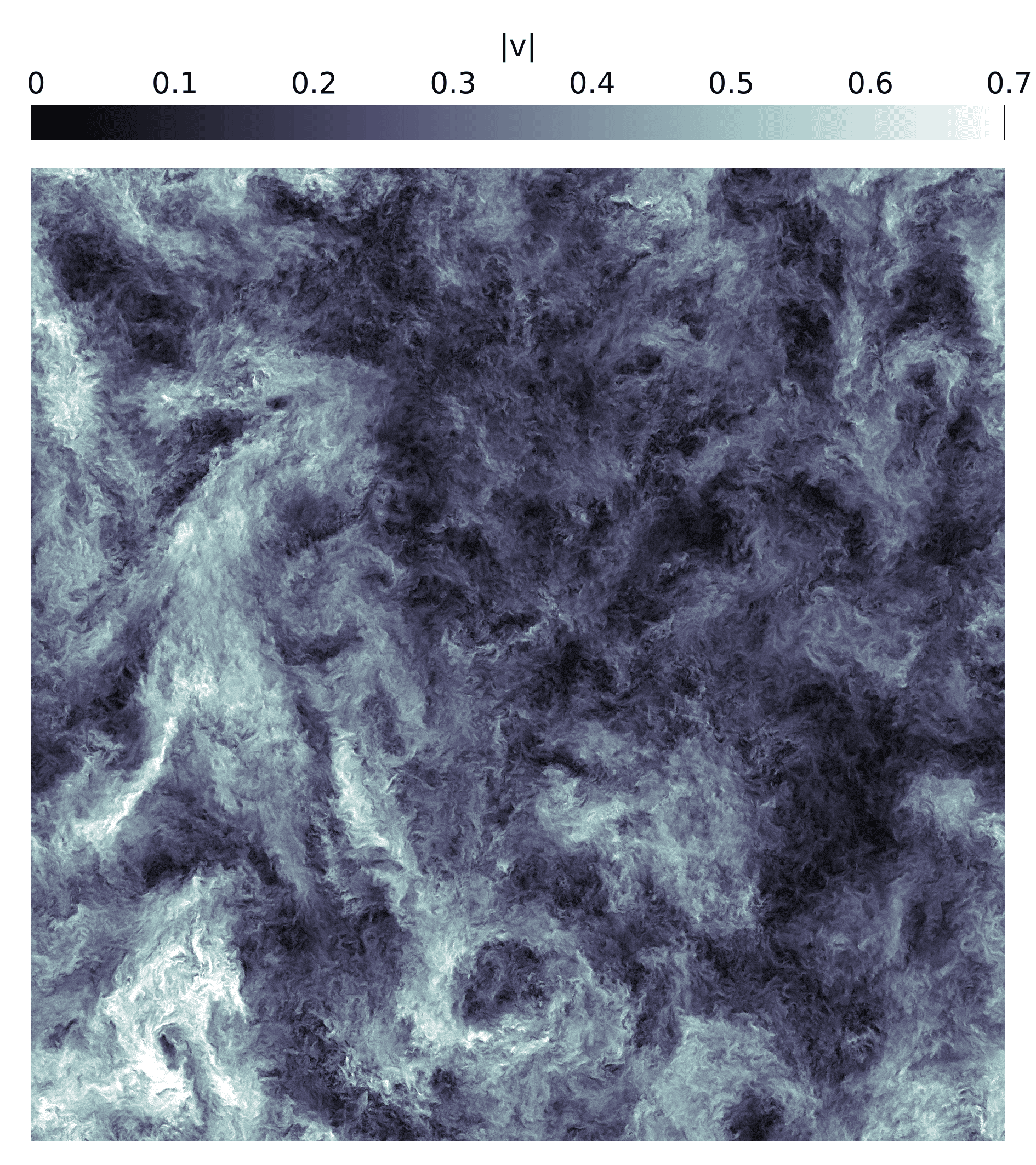}
    \caption{Velocity magnitude $|\mathbf{v}|$ at $t=10$ for the $3200^3$ simulation, shown in a slice of thickness $2h$ centered on $z=0$.}
    \label{fig:velocity}
\end{figure}

\begin{figure}
    \centering
    \includegraphics[width=\linewidth]{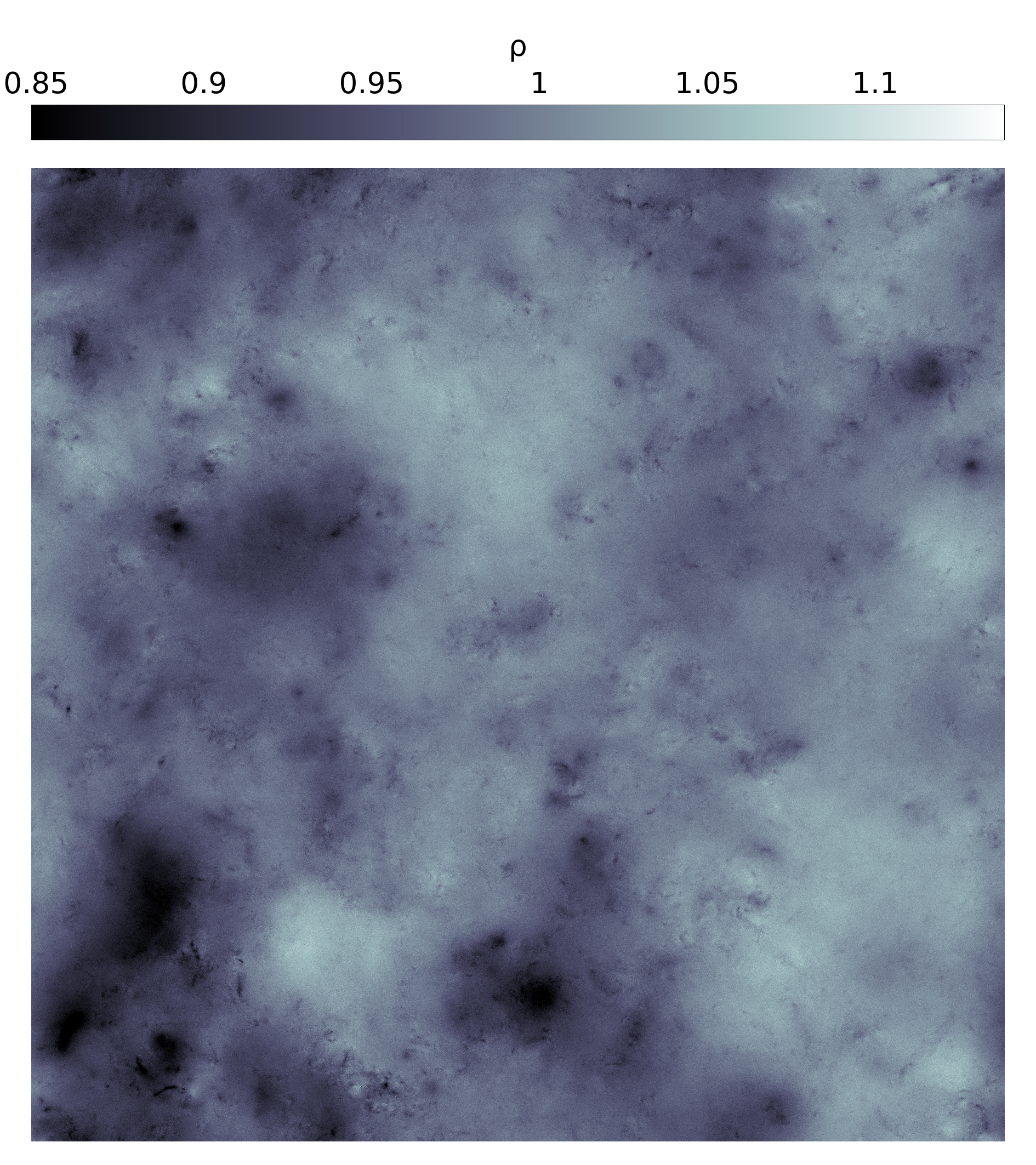}
    \caption{Density at $t=10$ for the $3200^3$ simulation, shown in a slice of thickness $2h$ centered on $z=0$.}
    \label{fig:density}
\end{figure}

\begin{figure}
    \centering
    \includegraphics[width=\linewidth]{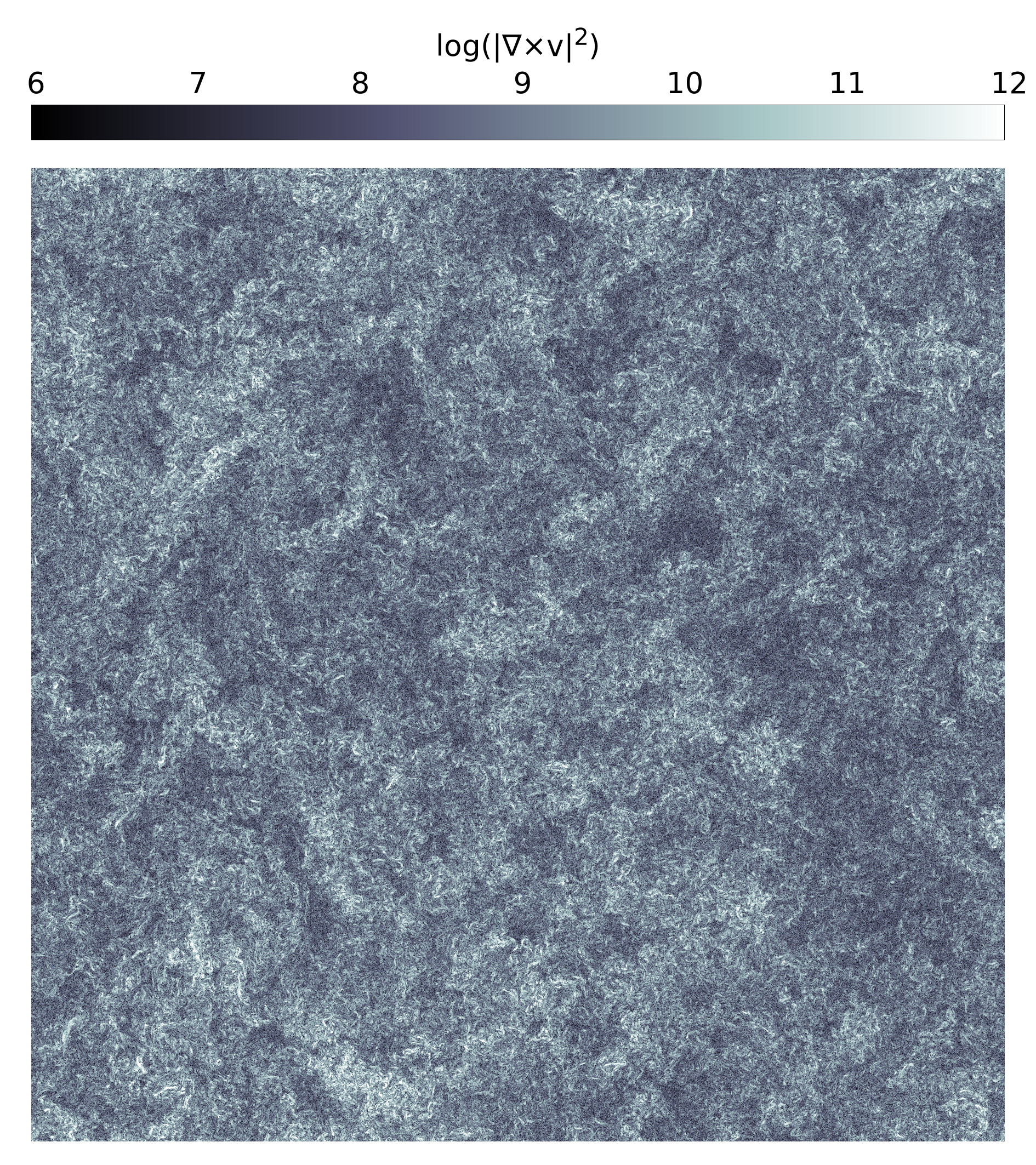}
    \caption{Enstrophy at $t=10$ for the $3200^3$ simulation, shown in a slice of thickness $2h$ centered on $z=0$.}
    \label{fig:enstrophy}
\end{figure}

Figures~\ref{fig:velocity} to \ref{fig:enstrophy} show a detailed image of a slice of the velocity magnitude, density, and enstrophy for the $3200^3$ simulation, respectively, where we can appreciate the very small-scale features resolved with SPH-EXA. All slices have a thickness of $2h$ and they are centered on $z=0$. Particles are displayed directly (no particle-to-grid interpolation is applied) and each slice contains $\sim 5.84\times 10^7$ particles.

\end{appendix}

\end{document}